\documentclass[runningheads]{svmult}

\usepackage{makeidx}   
\usepackage{graphicx}  
\usepackage{subeqnar}  
\usepackage{multicol}  
\usepackage{physprbb}  
\makeindex             

\begin{document}
\title*{From First Galaxies to QSOs\protect\newline Feeding the baby monsters}
%
%
%
%
\titlerunning{Feeding the baby monsters}
%
\author{L. Danese\inst{1} \and F. Shankar\inst{1} \and G.L. Granato\inst{2,1} \and
L. Silva\inst{3,1} \and A. Bressan\inst{2,1} \and G. De
Zotti\inst{2,1} \and P. Salucci\inst{1} \and M. Cirasuolo\inst{1}}
\authorrunning{Luigi Danese et al.}
%
%
\institute{SISSA/ISAS, via Beirut 2, I-34014 Trieste, Italy \and
INAF - Osservatorio Astronomico di Padova, I-35122 Padova, Italy
\and INAF - Osservatorio Astronomico di Trieste, I-34131 Trieste,
Italy}

\maketitle              

\begin{abstract}
We present a physical model for the coevolution of massive
spheroidal galaxies and active nuclei at their centers. Supernova
heating is increasingly effective in slowing down the star
formation and in driving gas outflows in smaller and smaller dark
matter halos. Thus the more massive protogalaxies virializing at
early times are the sites of faster star formation. The
correspondingly higher radiation drag causes a faster angular
momentum loss by the gas and induces a larger accretion rate onto
the central black hole. In turn, the kinetic energy of the
outflows powered by the active nuclei can unbind the residual gas
in a time shorter for larger halos. The model accounts for a broad
variety of dynamical, photometric and metallicity properties of
early-type galaxies, for the $M_{\rm BH}$--$\sigma$ relation and
for the local supermassive black-hole mass function.
\end{abstract}

\section{Introduction}
There is growing evidence that massive galaxies at high redshifts
are far more numerous than predicted by standard semi-analytic
models \cite{{Blain},{Scott},{Daddi},{Tezca},{Franx},{Dokkum}}.
Also, the [$\alpha/Fe$]-magnitude relation points towards a higher
abundance of $\alpha$-elements for more luminous/massive galaxies,
indicating that the time for type Ia SNe to enrich the ISM must
have been shorter for the more massive systems \cite{Matteucci}.
Extremely massive black holes (BH), with $\log \ (M_{\rm
BH}/M_{\odot})> 8$--9, must also have formed very quickly at early
cosmic times to power highly luminous quasars at redshift of up to
$> 6$  \cite{Fan}.

As stressed by \cite{{Granato01},{Granato04},{Tezca}}, in the
framework of the hierarchical clustering paradigm there are enough
massive dark halos to accommodate the observed high redshift
galaxies. The problem is to find a mechanism explaining the faster
and more efficient star-formation in more massive galactic halos.

The tight relationship between dynamic and photometric properties
of galactic bulges and the masses of BHs at their centers
\cite{{Ferrarese00},{Gebhardt},{McLure},{Tremaine},{Marconi}}
indicate that a key ingredient in this context is likely to be the
mutual feedback from star formation and the growing active
nucleus. Indeed the close interplay of the two components has a
crucial role in the {\it Anti-hierarchical Baryon Collapse} (ABC)
model by Granato et al. \cite{{Granato01},{Granato04}}, that
appears to overcome the main shortcomings of current semi-analytic
models. In the following we briefly describe this model and
summarize some of its predictions.

\section{The Model}
The ABC model applies to massive spheroidal galaxies and galactic
bulges (halo mass $\log(M_{\rm halo}/M_\odot)> 11.4$), virializing
at $z \ge 1.5$. In practice, it is assumed that massive halos
virializing in this redshift range end up as spheroidal galaxies,
while those virializing at later times host disks or irregular
galaxies. The virialization rate of these objects is given by the
positive term of the derivative of the Press \& Schechter
\cite{P&S} mass function, while the negative part (corresponding
to their disappearance due to merging) is negligible. Numerical
simulations \cite{{Wechsler},{Zhao}} have shown that the build-up
of DM halos consists of an early phase of fast accretion, during
which there is a rapid increase of the specific binding energy and
of the central potential well, and of a late phase of slow
accretion with no significant change of the binding energy and of
the circular velocity.

The diffuse gas within the DM well, shock heated to the virial
temperature of the halo, falls into the star forming regions at a
rate ruled by the cooling and dynamic timescales. The cooled gas
feels the feedback from SNe and from the central AGN which heat
and possibly expel the gas from the potential well. In small halos
a few SNe are sufficient to quench the star formation, while in
the big ones nothing prevents a huge starburst (1000
$M_{\odot}$/yr over 0.5 Gyr). Furthermore, the radiation drag
damps down the angular momentum of the cool gas \cite{Kawakatu},
letting it inflow into a reservoir around the central
super-massive BH (SMBH). Viscous drag then causes the gas to flow
from the reservoir into the SMBH, increasing its mass and powering
the nuclear activity until its feedback is strong enough to unbind
the residual ISM, thus stopping the star formation and letting the
active nucleus shine unobscured. The time required to sweep out
the ISM decreases with increasing halo mass, thus accounting for
the [$\alpha/Fe$]-magnitude relation. An almost passive evolution
of the stellar population, with a dormant SMBH, follows.

\begin{figure}
\begin{center}
\includegraphics[height=7truecm,width=.8\textwidth]{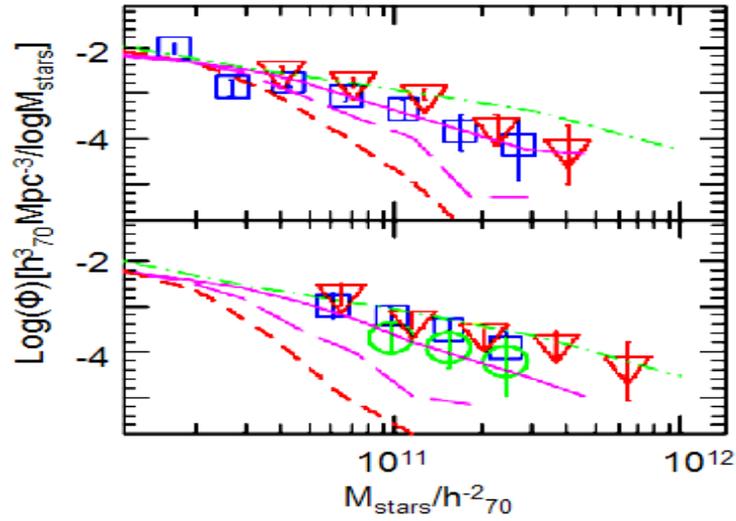}
\end{center}
\caption[]{Stellar mass functions of galaxies  in the ranges $1.
\leq z < 1.5$ (upper panel) and $1.5 \leq z < 2$ (lower panel)
derived by \cite{Fontana} (different symbols correspond to
different methods to estimate the stellar mass), compared with
theoretical predictions (dot-dashed line: ABC model; dashed and
thick solid lines: \cite{{Somerville04a},{Somerville04b}}; short
dashed line: \cite{Cole})} \label{FONTANA}
\end{figure}

\begin{figure}
\begin{center}
\includegraphics[height=6truecm,width=.8\textwidth]{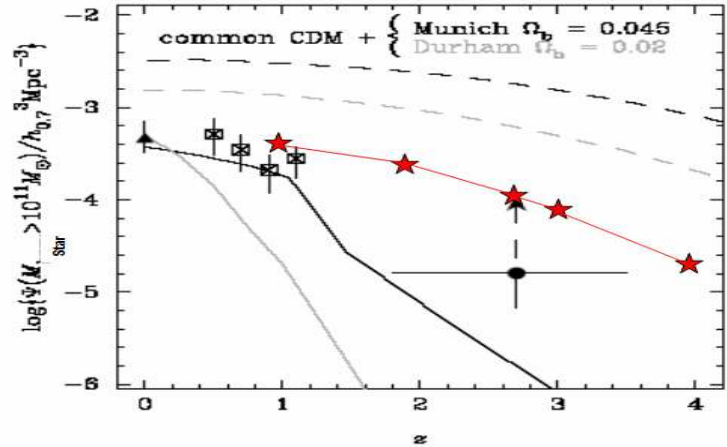}
\end{center}
\caption[]{Comoving number densities of galaxies with baryonic
masses $\geq 10^{11} M_{\odot}$ as a function of redshift. The
triangle and open rectangles show densities of massive stellar
systems at $z$=0 \cite{Cole1} and $z \sim 1$ \cite{Drory}. The
circle with upward arrow is the lower limit by \cite{Tezca}. The
solid curves show the predictions of \cite{{Kauffmann}} (upper)
and \cite{{Baugh}}; the dashed curves show the number densities of
halos with available baryonic masses $\geq 10^{11} M_{\odot}$ for
the values of $\Omega_b$ adopted by \cite{{Kauffmann}} (upper) and
\cite{{Baugh}}. The line with stars is the ABC model prediction.
Adapted from \cite{Tezca} } \label{TEZCA}
\end{figure}

\section{Results}
As shown by \cite{{Granato04}}, the ABC model, coupled with the
spectro-photometric code GRASIL by Silva et al. \cite{Silva},
accounts for a broad variety of data, including the SCUBA counts
at 850 $\mu$m (which are strongly under-predicted by the other
semi-analytic models), the corresponding preliminary redshift
distribution, and the local K-band luminosity function of massive
spheroidal galaxies. In Fig.~ \ref{FONTANA} we compare the model
predictions with the distribution of stellar masses in galaxies up
to $z\simeq 2$ determined by \cite{Fontana}. At $z \sim 3$ the
model predicts a comoving number density of galaxies with masses
$M_{star} \ge 10^{11} M_{\odot}$ of $n \sim 10^{-4}/\hbox{Mpc}^3$
in excellent agreement with the estimate by \cite{Tezca} (Fig.
\ref{TEZCA}).

\begin{figure}
\begin{center}
\includegraphics[height=5truecm,width=.7\textwidth]{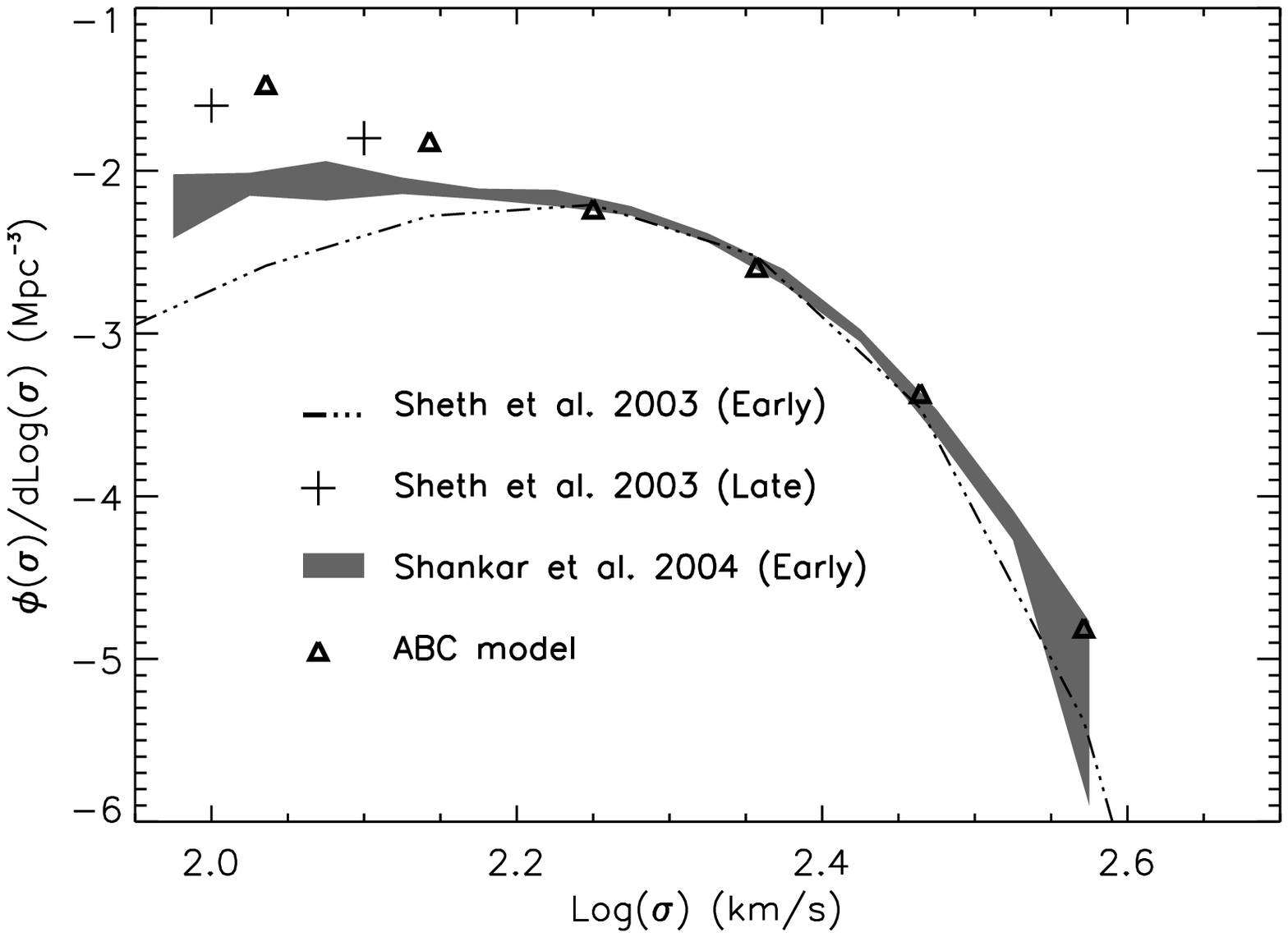}
\end{center}
\caption[]{Local velocity dispersion function derived from the
SDSS data \cite{{Sheth},{Shankar}} compared with that predicted by
the ABC model } \label{VDF}
\end{figure}

\begin{figure}
\begin{center}
\includegraphics[height=5.5truecm,width=.7\textwidth]{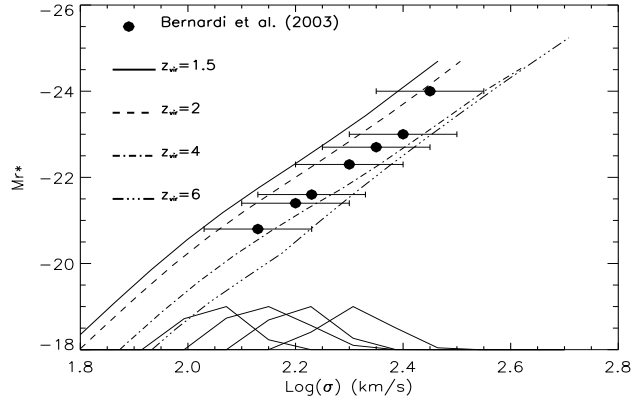}
\end{center}
\caption[]{Faber-Jackson relation predicted by the ABC model for
various virialization redshifts compared with data by
\cite{Bernardi}. Just above the $x$-axis are the normalized
distributions of velocity dispersions of galaxies in 4 absolute
magnitude bins 0.5 mag wide, centered at $Mr*=-20.2$, $-21$,
$-22$, and $-23$ (from left to right), as predicted by the ABC
model. The FWHMs of the distributions are remarkably close to the
observed values ($\hbox{FWHM} \sim 0.09$, see \cite{Bernardi}) }
\label{Lsigma}
\end{figure}

\begin{figure}
\begin{center}
\includegraphics[height=5truecm,width=.7\textwidth]{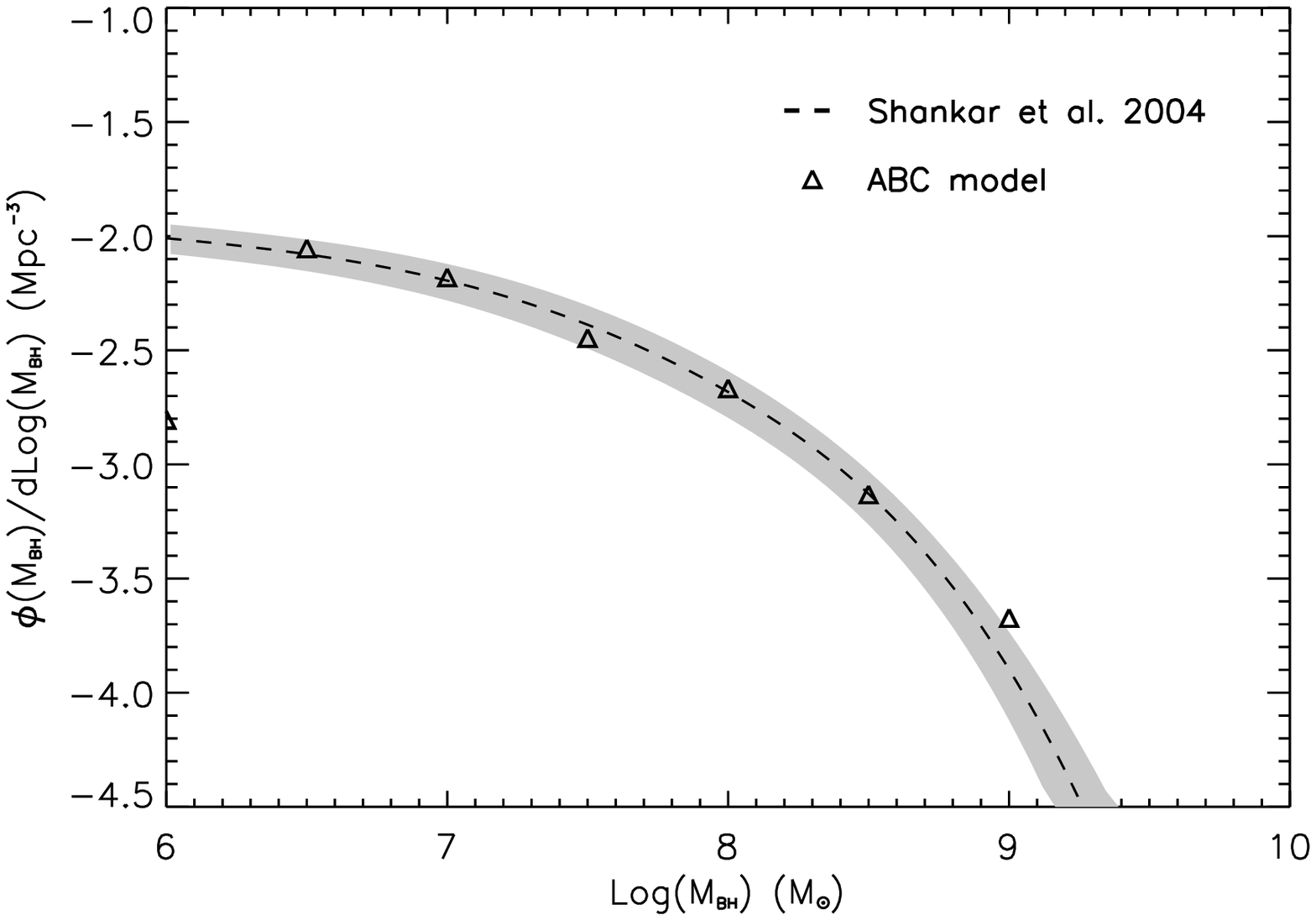}
\end{center}
\caption[]{Predicted local supermassive black hole mass function
compared with the one obtained by \cite{Shankar} from kinematic
and photometric data on spheroidal galaxies } \label{SMBH}
\end{figure}

For a given cosmology, a halo of mass $M_{\rm halo}$ virializing
at $z_{\rm vir}$ can be characterized by its circular velocity,
$V_{\rm halo}$. Thus, the velocity functions of halos virializing
at any given redshift, predicted by the standard hierarchical
clustering model for galaxy formation, can be straightforwardly
computed from their mass function. Integrating over redshift, for
$z_{\rm vir}\ge 1.5$, the local velocity function of spheroids can
be obtained. Adopting a constant ratio of the velocity dispersion
at $r_e/8$, $\sigma$, to $V_{\rm halo}$, $\sigma/V_{\rm
halo}=0.55$, consistent with the results by \cite{Ferrarese02},
Cirasuolo et al. \cite{Ciras} have found a very good match to the
velocity dispersion function (Fig. \ref{VDF}) of early-type
galaxies and bulges derived by \cite{{Sheth},{Shankar}} using SDSS
data \cite{Bernardi}. Furthermore, \cite{Ciras} have shown that
the relationship between the luminosity of spheroids and their
velocity dispersion, predicted by the ABC model, fits the
Faber-Jackson relation for NVSS galaxies \cite{Bernardi} (Fig.
\ref{Lsigma}); the observed distribution of velocity dispersions
at given luminosity is accounted for by the range of virialization
redshifts.

In the ABC model, the final (i.e. present day) mass of SMBHs in
galaxy centers is also a function of $M_{\rm halo}$ and $z_{\rm
vir}$ only. In Fig.~\ref{SMBH} the predicted local supermassive BH
mass function is compared with the recent estimate by
\cite{Shankar}. As shown by \cite{Granato04,Ciras} the model also
gives a good fit of the $M_{BH}-\sigma$ relation.


%


\section{Summary and conclusions}
The spheroid-SMBH coevolution model described here is based on the
idea of \emph{Anti-hierarchical growth} of the baryonic component
in DM halos. The heating from SNe is increasingly effective in
slowing down the star formation and driving gas outflows in
shallower potential wells. As a consequence, the star formation is
faster within the most massive halos. A higher star formation
implies a higher radiation drag, a faster SMBH fuelling and
growth, and therefore a stronger AGN kinetic output ($\propto
M_{BH}^{3/2}$, see \cite{Murray}) causing an earlier sweeping out
of the ISM. Thus the duration of the starburst and of the SMBH
growth is shorter for more massive halos ($\leq 1\,$Gyr for
$M_{\rm halo} \ge 10^{12} M_{\odot}$ and $3 \le z_{\rm vir} \le
6$).

During the intense starburst and SMBH growth phase the spheroid is
heavily dust obscured (SCUBA phase); the model indeed fits the
SCUBA counts and the (albeit limited) data on the redshift
distribution. We attribute the relatively high hard X-ray emission
recently detected from SCUBA galaxies by \cite{Ale} ($L_{X} \sim
10^{43}$--$10^{44}$ erg/s) as associated to the growing phase of
the central SMBH.


When the AGN reaches its maximum power, the ISM is blown away and
the AGN shines unobscured (QSO phase). The gas surrounding
powerful high-$z$ QSOs is therefore expected to exhibit at least
solar abundances and $\alpha$-enhance\-ment, as indeed found by
\cite{D'Odorico}.

Afterwards the spheroids evolve passively (ERO phase). The model
reproduces the mass and redshift distributions of sources detected
by deep K-band surveys, which proved to be extremely challenging
for all the other semi-analytical models, as well as the
dynamical/photometric properties of spheroidal galaxies in the
SDSS survey \cite{Bernardi}.

Finally, the model accounts for the local SMBH mass function for
$M_{\rm BH} > 10^7\,M_\odot$, and for the the $M_{\rm
BH}$--$\sigma$ relation. It predicts a steepening at $\sigma \le
150\,$km/s as the SMBH growth is hindered by the combined effect
of SNe heating and decreased radiation drag.

%

\end{document}